# Auditory Attention Decoding from Ear-EEG Signals: A Dataset with Dynamic Attention Switching and Rigorous Cross-Validation

Yuanming Zhang, Zeyan Song, Jing Lu, Fei Chen, Zhibin Lin

*Abstract*— Recent promising results in auditory attention decoding (AAD) using scalp electroencephalography (EEG) have motivated the exploration of cEEGrid, a flexible and portable ear-EEG system. While prior cEEGrid-based studies have confirmed the feasibility of AAD, they often neglect the dynamic nature of attentional states in real-world contexts. To address this gap, a novel cEEGrid dataset featuring three concurrent speakers distributed across three of five distinct spatial locations is introduced. The novel dataset is designed to probe attentional tracking and switching in realistic scenarios. Nested leave-one-out validation—an approach more rigorous than conventional single-loop leave-one-out validation—is employed to reduce biases stemming from EEG's intricate temporal dynamics. Four rule-based models are evaluated: Wiener filter (WF), canonical component analysis (CCA), common spatial pattern (CSP) and Riemannian Geometry-based classifier (RGC). With a 30-second decision window, WF and CCA models achieve decoding accuracies of 41.5% and 41.4%, respectively, while CSP and RGC models yield 37.8% and 37.6% accuracies using a 10-second window. Notably, both WF and CCA successfully track attentional state switches across all experimental tasks. Additionally, higher decoding accuracies are observed for electrodes positioned at the upper cEEGrid layout and near the listener's right ear. These findings underscore the utility of dynamic, ecologically valid paradigms and rigorous validation in advancing AAD research with cEEGrid.

†This work was supported by National Science Foundation of China with (Grant No. 12274221), Postgraduate Research & Practice Innovation Program of Jiangsu Province (Grant No. KYCX24_0141), and the National Key Research and Development Program of China (Grant No. 2023YFF1203502). (Corresponding author: *Zhibin Lin*.).

Yuanming Zhang, Zeyan Song, and Jing Lu are with Key Lab of Modern Acoustics, Nanjing University, Nanjing 210093, China, and also with the NJU-Horizon Intelligent Audio Lab, Horizon Robotics, Beijing 100094, China (e-mail: yuanming.zhang@smail.nju.edu.cn; lujing@nju.edu.cn).

Fei Chen is with the Department of Electrical and Electronic Engineering, Southern University of Science and Technology, Shenzhen 518055, China (e-mail: fchen@sustech.edu.cn).

Zhibin Lin is with Key Lab of Modern Acoustics, Nanjing University, Nanjing 210093, China (e-mail: zblin@nju.edu.cn).

Color versions of one or more of the figures in this article are available online at http://ieeexplore.ieee.org

The EEG recording experiments were approved by the Institutional Review Board of Nanjing University. Approval Number: OAP20240326001.

## I. INTRODUCTION

Individuals with normal hearing can effectively isolate and attend to a target speaker amid competing speech and environmental noise [1], [2], [3], [4]. In contrast, those with hearing impairment (HI) frequently struggle to focus on a target speaker, particularly in noisy contexts [5], [6]. Although hearing aids and cochlear implants substantially improve the quality of life for HI individuals, they remain limited in their capacity to steer auditory attention toward a specific talker based on cognitive intent. This limitation has motivated increasing interest in brain-computer interfaces (BCIs) and methods to decode auditory attentional states from neural signals [5], [7], [8], [9], [10], [11]. Over the past decade, auditory attention decoding (AAD)—a technique to identify a listener's attentional focus from neural signals—has garnered substantial interest following demonstrations of its feasibility [12], [13], [14], [15]. Among neuroimaging modalities, electroencephalography (EEG) is widely preferred for its non-invasiveness and practicality.

Two primary pipelines for decoding auditory attention from EEG signals have been proposed: stimulus reconstruction and directional focus decoding. Stimulus reconstruction entails estimating the attended speaker's audio features (e.g., envelope or mel spectrogram) from the listener's EEG signals, which are then compared to those of competing speakers to identify the attended talker [7], [16], [17]. Conversely, directional focus decoding—also termed selective AAD—seeks to directly predict the attended direction (e.g., left vs. right) from EEG signals [5], [13], [18], [19].

Both linear and nonlinear models are employed in these pipelines. Wiener filter (WF) model, which estimates the attended speech envelope using from time-lagged multichannel EEG signals, remains the most prevalent method for envelope prediction [9], [20]. For directional classification, common spatial pattern (CSP) and Riemannian geometry-based classifiers (RGC) have been developed, operating on log-energy features extracted from EEG signals [13], [18]. Although linear models primarily exploit second-order statistics of the training data, their inherent mismatch with the nonlinear dynamics of the auditory-brain pathway limits decoding performance [5], [9], [13], [14], [16], [18]. Nonetheless, their lower computational complexity enhances practicality for real-world applications.

To address these limitations, nonlinear approaches have emerged as promising alternatives. Deep neural networks (DNNs), particularly convolutional neural networks (CNNs), have enabled end-to-end prediction of attended direction directly from EEG signals [5], [19], [21], [22], [23]. Recently, Transformer-based architectures have also been adapted for BCI applications. For instance, the EEG-Deformer model integrates fine- and coarse-grained feature processing via a Conformer architecture, improving decoding across multiple BCI tasks [24], [25]. Similarly, M3ANet leverages Mamba—a state-space model alternative to Transformers—to directly separate attended speech from mixed audio using neural correlates, and SSM2Mel utilizes Mamba block to reconstruct the attended mel spectrogram from EEG signals [26], [27], [28], [29]. Graph neural networks (GNNs) have gained traction due to their natural alignment with EEG electrode topology, where electrodes are mapped to graph nodes [30], [31]. Despite these advancements, the application of DNNs to cEEGrid data remains underexplored [14], [32]. Ear-EEG signals, compared to scalp EEG, typically feature fewer channels and electrode placements farther from key brain regions, posing challenges for DNN models.

Prior studies have validated AAD feasibility with cEEGrid and other ear-EEG setups [14], [32], [33], [34]. However, they have largely overlooked the impact of dynamic attentional state switching on decoding performance [35], [36]. Existing studies have investigated the potential of decoding attentional switch from EEG signals, but no existing cEEGrid dataset captures EEG recordings with intra-trial attentional shifts, making the collection of comprehensive cEEGrid data essential for investigating AAD under dynamic attentional conditions [36], [37], [38].

Another critical challenge in BCI applications stems from unintended reliance on trial fingerprints. Specifically, a decoder (usually DNN models) can memorize trial fingerprints from EEG training segments. Later, the same fingerprints are extracted from test trials, and artificially inflates performance metrics are obtained when data are split using within-trial approaches [21], [23], [27], [39], [40], [41], [42]. Consequently, many researchers advocate for leave-one-trial-out or leave-one-subject-out cross-validation to obtain unbiased accuracy estimates in AAD tasks [25], [38], [43]. Notably, most existing models exhibit significant performance degradation under these rigorous validation schemes [38], [43]. However, the extent to which envelope reconstruction methods similarly degrade under such conditions remains uncharacterized.

The design of EEG experimental protocols significantly impacts decoding performance, particularly due to the presence of trial fingerprints. Notably, AAD decoders—especially nonlinear models—may exploit trial fingerprints to infer the experimental block from which an EEG segment originates. This risk is heightened when subjects attend to the same direction across consecutive trials within a block or follow predictable attentional patterns [5], [33], [44], [45], [46]. Mitigating trial bias through deliberate dataset design is therefore critical for developing robust AAD systems.

To address these challenges, a novel cEEGrid auditory attention dataset comprising 98 subjects is introduced [47], [48]. The dataset features three concurrent speakers symmetrically positioned in distinct spatial directions (left, front, and right). Participants were instructed to selectively attend to a designated target speaker. The experimental protocol was systematically randomized to minimize potential bias: attended directions were alternated randomly across consecutive trials. Competing speakers were partitioned into two non-overlapping groups to evaluate model generalization to unseen speakers. Furthermore, participants modulated their attentional focus through varied instructions, enabling dynamic tracking of attentional state switches within single-trial recordings.

Two linear models—Wiener Filter (WF) and Canonical Component Analysis (CCA)—and three DNN models—LSM-CNN, VLAAI, and SSM2Env—are evaluated on attended envelope reconstruction. Two rule-based models, namely common spatial pattern (CSP), and Riemannian geometry-based classifier (RGC) are evaluated for selective attention direction prediction. Comprehensive cross-validation is performed using three strategies: within-trial, standard leave-one-out, and nested leave-one-out. Leave-one-speaker-out experiments is further supplemented to assess generalization ability of these models on unseen speakers.

The contributions of this study are summarized as follows:

Firstly, a newly released cEEGrid auditory attention dataset comprising 98 participants is introduced. It is designed to enable investigation of dynamic attentional processes. To the best of our knowledge, this is the first large-scale (N=98) cEEGrid dataset for AAD that features three concurrent speakers in distinct spatial locations (left, front, and right) and employs a fully randomized attention-order protocol to minimize chronological biases.

Secondly, both statistics-based and DNN models are evaluated on this dataset and were fairly compared in different window lengths. The dataset and models are evaluated under three distinct cross-validation conditions: within-trial, leave-one-out, and nested leave-one-out. Additionally, model generalization ability to unseen speakers was assessed, addressing critical challenges posed by LRTC and speaker-specific biases.

Finally, Pearson correlation coefficient is used to analyze attentional state tracking and switching behavior within single EEG trials. Additionally, the feasibility of channel reduction is further explored to offer practical guidance for designing compact, miniaturized cEEGrid-based online devices.

The remainder of this paper is structured as follows. Section II details the three-speaker cEEGrid auditory attention dataset, including experimental protocols and EEG preprocessing pipelines. Section III briefly reviews the models employed, along with cross-validation experimental designs and result evaluation methodologies. Section IV presents experimental results and comparative discussion. Section V gives conclusion to this study.

## II. cEEGrid DATASET

### A. Experiment Setup

We present a novel, carefully designed, and minimally biased NJU auditory attention cEEGrid dataset [47], [48]. A total of 106 volunteers participated in the EEG signal collection experiments. All participants provided written informed consent prior to participation, with approval from the Nanjing University Ethics Committee (Approval No. OAP20240326001), and received financial compensation upon completion. Eight participants were excluded from further analysis due to hardware issues, yielding a final dataset of 98 subjects. All participants were students at Nanjing University. Gender and age were not considered or collected during recruitment. All self-reported normal hearing and mental health status.

For EEG acquisition, gold-plated cEEGrid electrodes, paired with the OpenBCI CytonDaisy Board, were used. One cEEGrid array was positioned behind participant's each ear, with the CytonDaisy Board recording 16 EEG channels simultaneously at a sampling rate of 125 Hz. Of the 10 channels per ear, 8 were used for data acquisition. An unused electrode on the left ear served as the reference channel (Figure 1-a).

Experimental instructions were displayed on a screen in front of participants. A small photodiode was placed at the corner of the screen to obtain trigger signals, which were used to synchronize audio stimuli with corresponding EEG recordings.

### B. Experiment procedure

Each participant completed 63 experimental trials, each lasting 30 seconds, resulting in 31.5 minutes of total EEG recording per participant. During each trial, participants were instructed to follow an on-screen arrow to identify their target speaker and maintain focus until an attentional switch cue—presented uniformly at random between the 13 and 17 seconds—prompted them to redirect attention. A brief training session preceded the experiments to familiarize participants with the task procedure, and regular breaks were taken during experiments to minimize fatigue. Following each trial, participants answered two single-choice comprehension questions: one based on the first 10 seconds of the attended speaker's stimuli and the other concerning the last 10 seconds.

The 63 trials were structured into seven distinct attentional switch tasks (see Figure 1 and Table 1), with nine trials per task. Each task's nine trials were subdivided into three groups of three trials. Two groups employed competing stimuli from speakers 1–3, while the third group used stimuli from speakers 4–6. Speaker 1 and Speaker 4 are female, and other speakers are male. Audio content comprised professionally narrated news segments originally broadcast by China Central Television, ensuring high-quality, naturalistic speech signals. For each group, audio stimuli were randomly selected without repetition from each speaker's audio library. Within a group, the three trials shared identical audio content but varied in the attended speaker and spatial direction. Trial order was fully randomized to prevent memorization of interfering stimuli, and each participant received a unique combination of attended and competing audio streams.

This rigorous design minimized consecutive trial repetitions:

**Table 1** Task instructions to the participants during EEG experiments.

| Task | Instruction |
| --- | --- |
| 1 | Attend to the specified speaker. Do not switch your attention. |
| 2 | Attend to the specified speaker. Switch your attention to the specified speaker when instructed. |
| 3 | Attend to the specified speaker. Switch your attention to another speaker as you wish when instructed. |
| 4 | Attend to the specified speaker. Ignore all speakers when instructed. |
| 5 | Attend to the specified speaker. Switch your attention to the speaker that would appear when instructed. |
| 6 | Attend to the specified speaker. Switch your attention to the remaining speaker when your attended speaker becomes silent. (Two speakers only.) |
| 7 | Attend to the specified speaker. Do not switch your attention. Ignore the interfering speaker that would appear. |

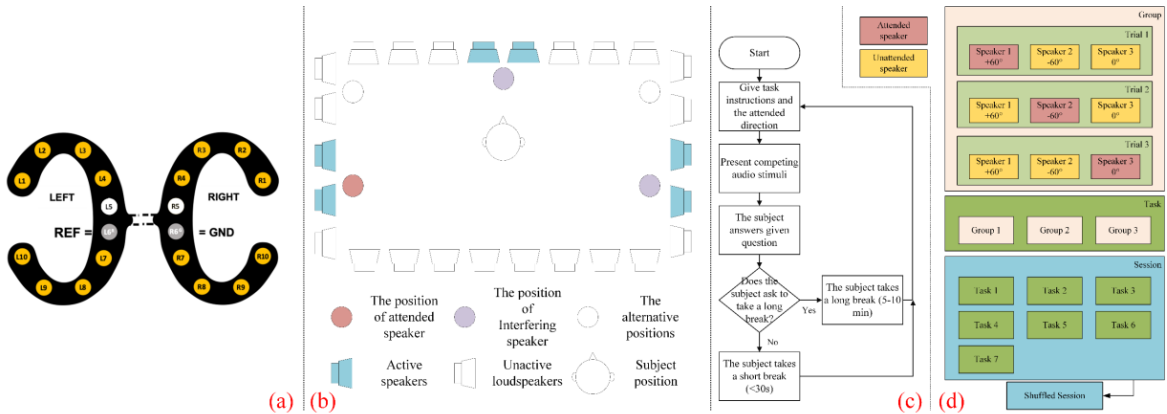

**Figure 1 A schematic diagram of the electrode placement and experiment scheme.** (a): The cEEGrid electrode placement. Yellow electrodes are used for further analysis. Gray electrodes are reference and ground electrodes. White electrodes are not included in this study. (b): A schematic diagram of the audio reproduction system. (c): A schematic diagram of the experiment procedure. (d): A schematic diagram of the experimental protocol. Each group consists of three trials using the same speakers and audio stimuli. Each task comprises three groups: two groups involve speakers 1, 2, and 3, while the remaining group includes speakers 4, 5, and 6. A single session consists of seven distinct tasks. The presentation order of trials is fully randomized at the session level.

participants were highly unlikely to (1) attend to the same spatial direction, (2) focus on the same speaker, or (3) perform the same task in adjacent trials. This protocol effectively eliminated overlapping attentional labels between consecutive trials, enhancing the robustness of AAD model evaluation. Schematic representations of the experimental details are provided in Figure 1 and Table 1.

Stimuli were presented via Genelec 8010 loudspeakers positioned within a 2 m × 1.5 m array. For each trial, the two loudspeakers closest to the target direction (±120°, ±60°, or 0°) were activated.

### C. EEG Signal Preprocessing

cEEGrid signals were aligned with audio stimuli using the rising edge of trigger signals. No additional alignment between EEG and audio signals was performed. EEG signals were first re-referenced to the dedicated reference channel. A preprocessing pipeline was then applied, including: (1) an 8th-order Butterworth infinite impulse response filter with a passband of [0.5 Hz, 62 Hz] to remove direct current drift and low-frequency noise; (2) a notch filter with a stopband of [48 Hz, 52 Hz] to eliminate line noise interference. Signal quality was further enhanced using artifact subspace reconstruction (ASR) [49], [50]. Independent Component Analysis (ICA) was performed on concatenated trial data for each subject. ICLabel, a component classification tool, was employed with an 80% confidence threshold to identify and remove artifact components (muscle, eye, and cardiac activity) [51]. Excluded channels were restored via interpolation.

### D. Envelope computation

A gammatone filter bank spanning 50 Hz to 5,000 Hz with 17 sub-bands, is applied to the audio signal. The absolute value of the filtered output is computed and then compressed by raising it to the power of 0.6. Finally, the sub-band signals are summed up and down-sampled to match the sampling rate of the EEG signals.

## III. METHODS

### A. Linear Wiener Filter

The linear WF backward regression model aims to minimize the Mean Squared Error (MSE) between the target and predicted signal. It applies a set of Finite Impulse Response (FIR) filters to the EEG signals to estimate the attended audio envelope and compares the Pearson's correlation coefficient (PCC) between the estimated envelope and both the attended and unattended envelopes.

The estimated signal can be written as

$$\hat{y}_a(t) = \sum_{c=0}^{C-1}\sum_{l=0}^{L-1} x_c(t-l)w_c(l) \quad (1),$$

where $\hat{y}_a$ is the estimated attended envelope, $x_c(t)$ is the $c$-th channel EEG signal, and $w_c$ represents the filter weights of the $c$-th channel filter with length $L$.

An equivalent expression of this equation in matrix form is

$$\hat{\mathbf{y}}_a = [\hat{y}_a(0);...;\hat{y}_a(T-1)]$$
$$= \mathbf{X}_{lag}\mathbf{w} = [\mathbf{X}_{0,lag};...;\mathbf{X}_{C-1,lag}][\mathbf{w}_0;...;\mathbf{w}_{C-1}] \in \mathbb{R}^{T\times 1} \quad (2),$$

where $\mathbf{X}_{c,lag}, \mathbf{w}_c$ are the matrix form of the lagged EEG signal and the filter weights.

$$\mathbf{X}_{lag} = [\mathbf{X}_{0,lag};...;\mathbf{X}_{C-1,lag}] \in \mathbb{R}^{T\times LC}$$

$$\mathbf{X}_{c,lag} = \begin{bmatrix} x_c(0) & \cdots & 0 \\ \vdots & & \vdots \\ x_c(t) & x_c(t-l) & x_c(t-L+1) \\ \vdots & & \vdots \\ x_c(T-1) & \cdots & x_c(T-L) \end{bmatrix} \quad (3).$$

$$\mathbf{w} = [w_0(0);...;w_{C-1}(0);...;w_{C-1}(L-1)] \in \mathbb{R}^{LC\times 1}$$

The filter weights $\mathbf{w}$ can be estimated by minimizing the MSE with an additional L$_2$ regularization term:

$$\mathbf{w} = \arg\min_{\mathbf{w}} \|\hat{\mathbf{y}}_a - \mathbf{y}_a\|^2 + \arg\min_{\mathbf{w}} \lambda\|\mathbf{w}\|^2 \quad (4).$$

The solution of this optimization problem is

$$\mathbf{w} = (\mathbf{X}_{lag}^T\mathbf{X}_{lag} + \lambda\mathbf{I})^{-1}\mathbf{X}_{lag}^T\mathbf{y}_a = (\mathbf{R}_{xx} + \lambda\mathbf{I})^{-1}\mathbf{r}_{xy} \quad (5),$$

where $\mathbf{R}_{xx}$ and $\mathbf{r}_{xy}$ are the *unnormalized* auto-correlation matrix and the cross-correlation vector.

### B. Canonical Correlation Analysis

The CCA model maximizes the correlation between the projection of the EEG and audio signals. Two different sets of FIR filters are applied to both EEG and audio signals to obtain their projections:

$$\bar{x}(t) = \sum_{c=0}^{C-1}\sum_{l=0}^{L_x-1} x_c(t-l)w_{x,c}(l)$$
$$\bar{y}_a(t) = \sum_{l=0}^{L_y} y_a(t-l)w_y(l) \quad (6).$$

An equivalent matrix form of Eq. (6) is

$$\bar{\mathbf{x}} = \mathbf{X}_{lag}\mathbf{w}_x, \bar{\mathbf{y}}_a = \mathbf{Y}_{lag}\mathbf{w}_y \quad (7).$$

To find a solution of the CCA model that maximizes the Pearson correlation coefficients (PCC) between $\bar{x}(t)$ and $\bar{y}_a(t)$, the singular value decomposition (SVD) of the whitened cross-covariance matrix between the EEG and attended audio signals is solved.

$$\mathbf{W}_x = \mathbf{R}_{xx}^{-1/2} \in \mathbb{R}^{L_xC\times L_xC}$$
$$\mathbf{W}_y = \mathbf{R}_{yy}^{-1/2} \in \mathbb{R}^{L_y\times L_y} \quad (8),$$
$$\mathbf{W}_x^T\mathbf{R}_{xy}\mathbf{W}_y = \mathbf{U}\mathbf{S}\mathbf{V}^T$$

where $\mathbf{R}_{xx}, \mathbf{R}_{yy}, \mathbf{R}_{xy}$ are the auto- and cross-covariance matrices of the EEG and audio signals, as defined in Eq. (5).

By extracting the first column of $\mathbf{U}, \mathbf{V}$, the components with the highest PCC between the projected vectors. And the second column corresponds to the projections with the second highest PCC are obtained. The first $N_{components}$ projection weights are computed as

$$\mathbf{w}_{x,i} = \mathbf{W}_x \mathbf{u}_i \in \mathbb{R}^{L_x C \times 1}$$
$$\mathbf{w}_{y,i} = \mathbf{W}_y \mathbf{v}_i \in \mathbb{R}^{L_y \times 1} \quad (9).$$
$$i = 0, 1, 2, ..., N_{components} - 1$$

### C. Common Spatial Pattern Classifier

The CSP Classifier applies different linear transformations to multichannel EEG signals to extract maximally discriminative features across different classes. The multiclass filter bank CSP (MC-FB-CSP) algorithm is described in this section.

A set of linear transformations is applied to the EEG signals

$$\mathbf{W}_{j,k} \in \mathbb{R}^{C \times F}, j = 0,1,...,N_{band}, k = 0,1,...,N_{class}$$
$$\hat{\mathbf{X}}_{j,k}(n) = \mathbf{X}_{j,k}(n)\mathbf{W}_{j,k} \in \mathbb{R}^{T \times F} \quad (10),$$

where $\mathbf{X}_{j,k}(n)$ is an EEG segment with band $j$ and label $k$. $N_{band}$ is the number of bands. $F$ is the number of features per frequency band. $N_{class}$ is the number of classes. The optimization goal is to maximize the energy contrast between the specified class $k$ and all classes.

$$\mathbf{W}_{j,k} = \arg\max_{\mathbf{w}_{j,k}} \frac{\mathbb{E}\left[\mathbf{W}_{j,k}^T \mathbf{X}_k^T \mathbf{X}_k \mathbf{W}_{j,k}\right]}{\mathbb{E}\left[\sum_{k=0}^{N_{class}} \hat{\mathbf{X}}_{j,k}^T \hat{\mathbf{X}}_{j,k}\right]} \quad (11).$$

Replacing expectations with ensemble averages, we solve this via generalized eigenvalue decomposition:

$$\mathbf{R}_{xx,k} \mathbf{W}_{j,k} = \mathbf{\Lambda} \mathbf{R}_{xx} \mathbf{W}_{j,k} \quad (12),$$

where $\mathbf{R}_{xx,k}$ is the ensemble autocorrelation w. r. t. class $k$, and $\mathbf{R}_{xx}$ is the autocorrelation of all samples. The transformation weights are selected as those with the largest generalized eigenvalues. Transformation weights are applied to each sample, and the log energy to obtain the CSP features are computed. The transformed features across all sub-bands are further concatenated. These features are subsequently split into training, validation and test sets to train a classifier, such as linear discriminator (LDA) or support vector machine (SVM), and predict the labels of each evaluation samples.

### D. Riemannian Geometry-based Classifier

The RGC classifier utilizes the symmetric positive definite (SPD) property of the covariance matrix to obtain a more accurate and stable estimation of the covariance matrix. The Riemannian mean of the covariance matrix is estimated based on the training set. The covariance matrices of each EEG sample are projected onto the tangent space of the Riemannian mean of the estimated covariance matrix.

The log-Euclidean approximation of the Riemannian mean of the covariance matrix is

$$\mathbf{R}_\Re = \exp\left(\sum_{n=0}^{N_{sample}-1} \log \mathbf{R}(n) / N_{sample}\right) \in \mathbb{R}^{C \times C} \quad (13),$$

where $\mathbf{R}(n) = \mathbf{X}^T(n)\mathbf{X}(n) \in \mathbb{R}^{C \times C}$ is the covariance matrix of the $n$-th EEG segment. The covariance matrix of each sample is then mapped to the tangent space of the Riemannian mean.

$$\mathbf{T}_{\mathbf{R}_\Re}(n) = \log \mathbf{R}_\Re^{-1/2} \mathbf{X}^T\ n\ \mathbf{X}\ n\ \mathbf{R}_\Re^{-1/2} / (T-1) \quad (14).$$

The tangent space matrix is also symmetric due to the symmetric property of both the Riemannian mean matrix and the covariance matrix. The upper triangular part of the tangent space matrix is vectorized as the representative feature of a sample segment. These representative features of each segment are used to train and test an LDA model to obtain the prediction of the attended class.

### E. DNN Models

Three DNN models, namely LSM-CNN, VLAAI, and SSM2Env, are used to reconstruct the attended envelope [16], [19], [27]. It is noted that SSM2Mel is designed to restore the attended mel spectrum, and is adopted to predict envelope in this paper (SSM2Env). Both LSM-CNN and VLAAI comprise of stacked convolution blocks to capture EEG features, while LSM-CNN utilize the learnable spatial mapping (LSM) module to build an effective 2D representation of EEG electrodes. Instead, VLAAI and SSM2Env treat EEG electrodes as a single dimension. SSM2Env leverages the state space mamba module to bidirectionally extract effective information from EEG signals.

## IV. DATA PREPERATION & TRAINING

### A. Cross-Validation

Existing research has addressed the critical importance of leave-one-out cross-validation in EEG decoding studies [19], [25], [43], [43]. While the susceptibility of classification models to EEG's chronological biases is well-documented, the extent to which regression models—such as those predicting attended speech envelopes from neural signals—face similar challenges under rigorous cross-validation remains understudied [38]. To address this gap, the performance of both WF and CCA models are compared across multiple validation protocols.

To obtain unbiased experimental results, many studies advocate the use of leave-one-out experimental conditions. Specifically, the training, validation, and test sets are not split within-trial. Several studies have utilized a less constrained leave-one-out experiment, in which the model is directly tested immediately after training, and hyperparameter optimization is performed on the test set [14]. To clearly distinguish between these two experimental conditions, the latter is referred to as leave-one-out and the former, stricter one, as nested-leave-one-out. Figure 2 illustrates these protocols.

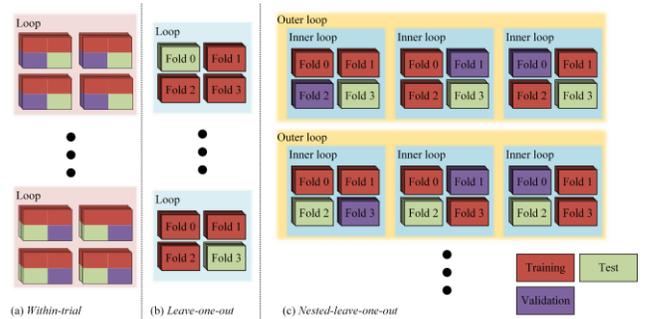

**Figure 2 A schematic diagram for the within-trial, leave-one-out, nested-leave-one-out cross-validation conditions.** In this figure, each rectangular solid box represents a trial. Stacked trials form a fold. (a) The *within-trial* cross-validation. Each trial is split into non-overlapping segments. These segments are then randomly split into training, validation and test sets. (b) The *leave-one-out* cross-validation. Trials are split into several folds, and each fold contains unique trials. These folds are then split into training and test sets. (c) The *nested-leave-one-out* cross-validation. Trials are split into several folds, and each fold also contains unique trials. In the inner loop of the cross-validation, the test fold remains unchanged, and the validation fold iterates over all other folds. The test fold is different in each outer loop. One outer loop contains several inner loops.

**Table 2 Decoding metrics of individualized models across several decision window lengths.** The hyperparameters of WF and CCA models are tuned for each individual to maximize the decoding accuracy of inner loops (validation set) with a decision window length of 30 s. The best hyperparameters of WF and CCA models are then selected and used to train and evaluate models on other conditions. The CSP and RGC models, along with DNN models, use empirically determined hyperparameters across all conditions. Results are evaluated on the test set and averaged across all test folds and subjects. The ΔPCC represents the difference between the attended and one unattended PCC. Decoding is marked as correct if both $\Delta PCC_1$ and $\Delta PCC_2$ is positive. The chance levels of accuracy and F1 are both 1/3.

| Metric | Hyper-parameter Search | Window Length | 2 s | 5 s | 10 s | 30 s |
|---|---|---|---|---|---|---|
| Accuracy | Individualized | WF | 0.347±0.035 | 0.365±0.057 | **0.378±0.086** | **0.415±0.129** |
| | | CCA | 0.349±0.037 | 0.364±0.062 | 0.376±0.090 | 0.414±0.132 |
| | Empirically Set | CSP | 0.379±0.122 | **0.377±0.132** | 0.371±0.150 | - |
| | | RGC | **0.382±0.125** | 0.363±0.131 | 0.375±0.148 | - |
| | | LSM-CNN | 0.336±0.002 | 0.342±0.006 | 0.344±0.012 | 0.326±0.013 |
| | | VLAAI | 0.331±0.002 | 0.337±0.006 | 0.334±0.014 | 0.318±0.019 |
| | | SSM2Env | 0.330±0.002 | 0.339±0.007 | 0.338±0.010 | 0.336±0.010 |
| F1 score | Individualized | WF | 0.478±0.039 | 0.496±0.063 | **0.509±0.093** | **0.547±0.131** |
| | | CCA | 0.479±0.042 | 0.495±0.068 | 0.508±0.097 | 0.543±0.134 |
| | Empirically Set | CSP | 0.147±0.057 | 0.171±0.065 | 0.195±0.085 | - |
| | | RGC | 0.149±0.056 | 0.149±0.056 | 0.196±0.082 | - |
| | | LSM-CNN | 0.491±0.002 | 0.499±0.007 | 0.500±0.014 | 0.480±0.015 |
| | | VLAAI | 0.485±0.002 | 0.493±0.007 | 0.489±0.016 | 0.470±0.021 |
| | | SSM2Env | **0.495±0.002** | **0.505±0.007** | 0.503±0.012 | 0.491±0.011 |
| Attended PCC | Individualized | WF | 0.012±0.019 | 0.013±0.019 | 0.015±0.020 | 0.016±0.019 |
| | | CCA | 0.014±0.028 | 0.018±0.032 | 0.020±0.033 | 0.022±0.031 |
| | Empirically Set | LSM-CNN | 0.032±0.001 | 0.036±0.001 | 0.045±0.002 | 0.070±0.001 |
| | | VLAAI | 0.032±0.002 | 0.035±0.001 | 0.046±0.002 | 0.075±0.010 |
| | | SSM2Env | **0.036±0.001** | **0.053±0.002** | **0.073±0.002** | **0.131±0.002** |
| Unattended $PCC_1$ | Individualized | WF | 0.002±0.014 | 0.002±0.014 | 0.003±0.014 | 0.003±0.012 |
| | | CCA | 0.005±0.020 | 0.006±0.024 | 0.006±0.027 | 0.007±0.024 |
| | Empirically Set | LSM-CNN | 0.028±0.001 | 0.028±0.001 | 0.040±0.002 | 0.071±0.002 |
| | | VLAAI | 0.032±0.002 | 0.029±0.001 | 0.043±0.001 | 0.075±0.010 |
| | | SSM2Env | 0.037±0.002 | 0.048±0.001 | 0.070±0.001 | 0.128±0.002 |
| Unattended $PCC_2$ | Individualized | WF | **0.002±0.014** | **0.002±0.013** | **0.002±0.013** | **0.002±0.012** |
| | | CCA | 0.004±0.020 | 0.005±0.024 | 0.006±0.026 | 0.008±0.024 |
| | Empirically Set | LSM-CNN | 0.030±0.002 | 0.033±0.002 | 0.041±0.002 | 0.071±0.002 |
| | | VLAAI | 0.034±0.002 | 0.034±0.002 | 0.044±0.002 | 0.075±0.010 |
| | | SSM2Env | 0.038±0.001 | 0.051±0.001 | 0.071±0.001 | 0.131±0.001 |
| $\Delta PCC_1$ | Individualized | WF | **0.010±0.020** | 0.011±0.020 | 0.012±0.021 | 0.013±0.019 |
| | | CCA | **0.010±0.025** | **0.013±0.026** | **0.014±0.028** | **0.016±0.027** |
| | Empirically Set | LSM-CNN | 0.003±0.001 | 0.007±0.001 | 0.005±0.002 | -0.001±0.002 |
| | | VLAAI | 0.000±0.002 | 0.006±0.001 | 0.003±0.002 | -0.002±0.002 |
| | | SSM2Env | -0.001±0.001 | 0.005±0.002 | 0.004±0.003 | 0.003±0.001 |
| $\Delta PCC_2$ | Individualized | WF | 0.009±0.020 | 0.011±0.021 | 0.013±0.021 | 0.013±0.020 |
| | | CCA | **0.010±0.025** | **0.013±0.027** | **0.014±0.028** | **0.015±0.027** |
| | Empirically Set | LSM-CNN | 0.002±0.002 | 0.002±0.003 | 0.004±0.003 | -0.001±0.002 |
| | | VLAAI | -0.001±0.001 | 0.001±0.003 | 0.002±0.003 | -0.002±0.003 |
| | | SSM2Env | -0.002±0.001 | 0.002±0.002 | 0.003±0.003 | -0.001±0.002 |

Additionally, additional experiments are conducted to evaluate the models' ability to generalize to unseen speakers. In this setting, the training set comprises trials with attended speakers 1–3, while validation and test sets use trials with speakers 4–6 (unseen during training). This experimental design enables us to examine whether the models rely on speaker-specific statistical regularities or capture generalizable attentional cues independent of speaker identity.

### B. Model training

All methodologies were implemented utilizing Python 3.12. The Python package Optuna was employed for hyperparameter tuning [52]. One hundred trials were conducted, and the Tree-structured Parzen Estimator (TPE) algorithm, which is the default search algorithm provided by Optuna, was utilized to identify optimal hyperparameters that maximize the decoding accuracy on the validation set [53], [54], [55], [56]. Per-channel z-score normalization was applied to both cEEGrid signals and the competing envelopes to enhance decoding performance. EEG signals and the envelopes were resampled to 40 Hz to reduce computational complexity while maintaining rigorous performance.

For DNN models, Superhuge, a python package built for AAD based on Pytorch and Pytorch Lightning, is used to load data, train models, and log results. AdamW with a learning rate of 5e-3 is used [57]. Models are trained for 100 epochs. ReduceLROnPlateau is used to dynamically control the learning rate based on validation loss. The negative Pearson correlation coefficient serves as the loss function, and early stopping is applied according to validation performance to prevent overfitting.

### C. Model performance evaluation

Pearson correlation coefficient (PCC) is used to evaluate the correlation between the reconstructed envelope and the attended or unattended envelope.

$$\rho_m = \frac{\mathbb{E}\left[\left(\hat{y}_a - \mathbb{E}\hat{y}_a\right)\left(y_m - \mathbb{E}y_m\right)\right]}{\sqrt{\mathbb{E}\left[\left(\hat{y}_a - \mathbb{E}\hat{y}_a\right)^2\right]\mathbb{E}\left[\left(y_m - \mathbb{E}y_m\right)^2\right]}} \quad (15).$$

$$m = 0, 1, ..., N_{speakers} - 1$$

By comparing the PCC values between the reconstructed stimuli and the competing envelopes, the attended speaker is determined as the one corresponding to the highest PCC value. Additionally, the accuracy and F1-score of attention decoding are evaluated. For classification models, the decoding accuracy and F1-score are used to assess the model's performance.

## V. RESULTS & DISCUSSION

### A. Overall decoding performance

Table 2 summarizes the decoding performance of the WF and CCA models with individually tuned hyperparameters, alongside the CSP and RGC models with empirically set hyperparameters. Both classification-based models (CSP and RGC) achieve higher decoding accuracy when the decision window length is 2 seconds or 5 seconds. Furthermore, when the decision window length is set to 10 seconds, the CSP and RGC models attain a decoding accuracy comparable to that of the WF and CCA models. However, both CSP and RGC models exhibit significantly deterioration of F1-score under this condition. Despite their reduced F1-score, the decoding performance of both CSP and RGC models improves with longer decision window length, aligning with previous findings [8], [9].

The regression models, WF and CCA, consistently exceed chance-level performance (i.e., 1/3) across all experimental conditions reported in Table 2. Both WF and CCA models achieve improved decoding accuracy, F1-score, and attended PCC values as the decision window length increases. However, the unattended PCC also slightly increases with longer input EEG segments, particularly in the case of CCA models. Accordingly, the difference in PCC values between the attended speaker and the two interfering speakers is also reported in Table 2. It is evident that the CCA model demonstrates consistently higher PCC differences across all decision window lengths, albeit with a higher standard deviation, which results in no significant difference in decoding accuracy between the CCA and WF models. One plausible explanation is that, owing to its enhanced modeling capacity, the CCA model reconstructs the target envelope by leveraging speech features common to both the attended and unattended speakers.

Notably, when provided with longer decision window durations, DNN-based models demonstrate a measurable improvement in their ability to correlate with the attended speaker's PCC metric. Despite this enhancement in attended PCC, critical evaluation reveals that both the overall decoding accuracy and the differential PCC (ΔPCC)—defined as the discrepancy between attended and unattended PCC values—remain statistically indistinguishable from chance levels across experimental conditions. These observations lead us to hypothesize that DNN architectures may be prioritizing the learning of generalizable, rather than attended-speaker-specific, neural patterns: by optimizing for broad correlation between reconstructed signals and speech envelopes, they inadvertently fail to isolate the neural signatures unique to the attended speaker. Specifically, rather than successfully decoding the true attended speech envelope from the EEG input data, these models appear to learn a superficial "predicted envelope" that exhibits strong, non-selective correlation with both attended and interfering speech envelopes. Given their inability to recover the target speaker's envelope, these DNN models are therefore excluded from subsequent analytical stages.

### B. The impact of cross-validation protocols

Table 3 and Table 4 summarize the decoding performance of the WF and CCA models across diverse cross-validation experiments and decision window lengths. As illustrated in Table 4, individualized hyperparameter tuning leads to higher decoding performance in leave-one-out experimental settings, but the performance declines in the more rigorous nested leave-one-out cross-validation experiments. This observation aligns with prior studies and emphasizes the significance of selecting an appropriate hyperparameter search strategy and cross-validation protocol [14].

**Table 3 Decoding performance of the WF and CCA models on different hyperparameter search and cross-validation experiments.** The decision window length is set to **2 seconds**. Group-wise hyperparameters tuned in 30 seconds are used. All models are trained and evaluated subject-by-subject. The ΔPCC represents the difference between the attended and one unattended PCC. Decoding is marked as correct if both ΔPCC$_1$ and ΔPCC$_2$ is positive. The chance levels of accuracy and F1 are both 1/3.

| Metric | Cross-validation | within-trial | leave-one-out | nested-leave-one-out |
|---|---|---|---|---|
| Accuracy | WF | 0.353±0.040 | 0.353±0.036 | 0.353±0.036 |
|  | CCA | 0.356±0.042 | 0.355±0.039 | 0.355±0.038 |
| F1 score | WF | 0.484±0.045 | 0.483±0.040 | 0.483±0.040 |
|  | CCA | 0.487±0.047 | 0.485±0.043 | 0.486±0.042 |
| Attended PCC | WF | 0.015±0.019 | 0.014±0.019 | 0.014±0.019 |
|  | CCA | 0.017±0.025 | 0.016±0.025 | 0.016±0.025 |
| Unattended PCC$_1$ | WF | 0.003±0.015 | 0.003±0.013 | 0.003±0.013 |
|  | CCA | 0.004±0.017 | 0.005±0.016 | 0.005±0.016 |
| Unattended PCC$_2$ | WF | 0.003±0.014 | 0.003±0.013 | 0.002±0.013 |
|  | CCA | 0.005±0.015 | 0.004±0.014 | 0.004±0.015 |
| ΔPCC$_1$ | WF | 0.012±0.020 | 0.011±0.019 | 0.011±0.019 |
|  | CCA | 0.013±0.020 | 0.011±0.019 | 0.011±0.019 |
| ΔPCC$_2$ | WF | 0.012±0.021 | 0.011±0.019 | 0.011±0.019 |
|  | CCA | 0.012±0.022 | 0.012±0.021 | 0.012±0.021 |

**Table 4 Decoding performance of the WF and CCA models on different hyperparameter search and cross-validation experiments.** The decision window length is set to **30 seconds**. Hyperparameter is tuned based on the validation set of each subject for individualized models, and is tuned on the validation set of all subjects for group-wise models. All models are trained and evaluated subject-by-subject. The ΔPCC represents the difference between the attended and one unattended PCC. Decoding is marked as correct if both $\Delta PCC_1$ and $\Delta PCC_2$ is positive. The chance levels of accuracy and F1 are both 1/3.

| Metric | Hyper-parameter search | Cross-validation | leave-one-out | nested-leave-one-out | leave-one-speaker-out | nested-leave-one-speaker-out |
|---|---|---|---|---|---|---|
| Accuracy | Individualized | WF | **0.575±0.086** | 0.415±0.129 | **0.689±0.112** | **0.415±0.210** |
| | | CCA | 0.571±0.091 | 0.414±0.132 | 0.685±0.115 | 0.402±0.201 |
| | Group | WF | **0.444±0.008** | **0.438±0.010** | **0.460±0.010** | **0.441±0.011** |
| | | CCA | 0.441±0.010 | 0.436±0.013 | 0.453±0.005 | 0.422±0.012 |
| F1 score | Individualized | WF | **0.702±0.073** | **0.547±0.131** | **0.790±0.088** | **0.537±0.223** |
| | | CCA | 0.698±0.079 | 0.543±0.134 | 0.786±0.092 | 0.525±0.221 |
| | Group | WF | **0.576±0.007** | **0.570±0.009** | **0.582±0.009** | **0.563±0.007** |
| | | CCA | 0.573±0.010 | 0.568±0.012 | 0.574±0.006 | 0.544±0.009 |
| Attended PCC | Individualized | WF | 0.023±0.017 | 0.016±0.019 | 0.025±0.017 | 0.015±0.023 |
| | | CCA | **0.033±0.033** | **0.022±0.031** | **0.037±0.040** | **0.020±0.049** |
| | Group | WF | 0.019±0.001 | 0.019±0.001 | 0.020±0.002 | 0.020±0.003 |
| | | CCA | **0.022±0.002** | **0.021±0.001** | **0.023±0.002** | **0.021±0.005** |
| Unattended $PCC_1$ | Individualized | WF | **0.000±0.010** | 0.003±0.012 | -0.003±0.014 | **0.002±0.020** |
| | | CCA | 0.003±0.027 | 0.007±0.024 | **-0.004±0.035** | 0.005±0.047 |
| | Group | WF | **0.004±0.000** | **0.004±0.001** | **0.004±0.001** | **0.004±0.001** |
| | | CCA | 0.006±0.001 | 0.007±0.001 | 0.005±0.001 | 0.005±0.002 |
| Unattended $PCC_2$ | Individualized | WF | **0.000±0.009** | **0.002±0.012** | -0.002±0.014 | **0.003±0.019** |
| | | CCA | 0.003±0.025 | 0.008±0.024 | **-0.004±0.035** | 0.010±0.047 |
| | Group | WF | **0.004±0.000** | **0.003±0.001** | **0.004±0.001** | **0.004±0.002** |
| | | CCA | 0.005±0.001 | 0.005±0.002 | 0.005±0.002 | 0.007±0.002 |
| $\Delta PCC_1$ | Individualized | WF | 0.022±0.014 | 0.013±0.019 | 0.028±0.019 | 0.013±0.028 |
| | | CCA | **0.030±0.020** | **0.016±0.027** | **0.040±0.030** | **0.015±0.041** |
| | Group | WF | **0.016±0.001** | **0.015±0.001** | 0.016±0.001 | **0.016±0.002** |
| | | CCA | 0.015±0.001 | 0.015±0.001 | **0.018±0.001** | 0.016±0.002 |
| $\Delta PCC_2$ | Individualized | WF | 0.023±0.015 | 0.013±0.020 | 0.027±0.019 | **0.012±0.029** |
| | | CCA | **0.030±0.021** | **0.015±0.027** | **0.037±0.029** | 0.011±0.044 |
| | Group | WF | 0.016±0.002 | 0.016±0.002 | 0.017±0.002 | **0.016±0.003** |
| | | CCA | 0.016±0.002 | 0.016±0.002 | **0.018±0.001** | 0.015±0.004 |

The leave-one-speaker-out cross-validation experiments are implemented following the identical procedural framework employed in the leave-one-trial-out experiments. Relative to the leave-one-trial-out condition, both WF and CCA models demonstrated significantly improved decoding performance within the leave-one-speaker-out scenario. Nevertheless, no significant between-scenario differences were observed in the results of either model when comparing nested leave-one-out and nested leave-one-speaker-out cross-validation approaches. These findings indicate that envelope reconstruction models exhibit limited susceptibility to speaker-specific features and possess the capacity to generalize effectively to previously unseen speakers. Nonetheless, due to inherent limitations in the EEG experimental protocol, the validation and test sets still contained overlapping speakers, which may have induced speaker-specific hyperparameter tuning and consequently resulted in overestimated decoding performance.

Given the limited duration of individual EEG trials, results from comparative experiments using a 2-second decision window length revealed no significant differences in performance between within-trial validation and other leave-one-out experimental conditions under the tested parameters (see Table 3). Owing to the constrained statistical power of short EEG segments and the sparse electrode configuration of cEEGrid systems, both WF and CCA models exhibited reduced decoding accuracy and lower reconstructed PCC values when employing a 2-second decision window length.

In contrast to classification models, no overestimation of decoding performance is observed. Nevertheless, it remains essential to conduct comprehensive investigations into whether performance overestimation occurs in other EEG datasets and experimental conditions to substantiate the conclusion that within-trial experiments for regression models do not yield overestimated decoding results. Additionally, nested leave-one-out experiments remain critical for obtaining unbiased results.

### C. Analysis of filter weights

To further examine the functional contributions of individual cEEGrid channels and assess the impact of channel optimization on model performance, both the maximum absolute values and the mean squared values of the filters associated with each cEEGrid channel are computed. Figure 3 depicts the channel-wise weight distributions of the WF and CCA models plays a critical role in decoding the attended envelope, no significant differences were observed between the combined weights of the eight left channels and those of the eight right channels in either model ($p > 0.05$). For the CCA model, statistical tests indicate no significant difference between the upper four channels and the lower four channels for both the left and right ears (p>0.05). In contrast, the WF model exhibited significant differences between the upper and lower channels across both ears. This observed disparity between upper and lower channel weights aligns with the anatomical observation that the upper four channels (L1–L4 and R1–R4) are positioned closer to cortical brain regions, particularly those involved in speech processing and attention mechanisms. Collectively, these findings suggest that the WF model may possess the capacity to extract the attended envelope using only the upper four channels of each ear.

**Table 5 Decoding performance of Wiener Filter model with partial cEEGrid channels.** The decision window length is set to 30 seconds. Group-wise tuned hyperparameters are chosen based on full electrode layout and are used for the resting electrode placements. Nested-leave-one-trial-out cross-validation is applied, and the results in test sets are reported. All models are trained and evaluated subject-by-subject. The ΔPCC represents the difference between the attended and one unattended PCC. Decoding is marked as correct if both $\Delta PCC_1$ and $\Delta PCC_2$ is positive. The chance levels of accuracy and F1 are both 1/3. Colored electrodes are used to train and evaluate the model, grayed-out electrodes are not used, except the ground and reference electrodes.

| Metric | Electrode layout | Full | Left | Right | Upper | Lower | Left-Upper | Right-Upper |
|---|---|---|---|---|---|---|---|---|
| Accuracy | | 0.438±0.010 | 0.414±0.135 | 0.424±0.140 | **0.446±0.151** | 0.389±0.131 | 0.404±0.137 | 0.423±0.142 |
| F1 score | | 0.570±0.009 | 0.547±0.143 | 0.557±0.146 | **0.578±0.154** | 0.520±0.143 | 0.537±0.146 | 0.555±0.150 |
| Attended PCC | | 0.019±0.001 | 0.013±0.016 | 0.017±0.020 | **0.021±0.021** | 0.009±0.014 | 0.012±0.016 | 0.017±0.020 |
| Unattended $PCC_1$ | | 0.004±0.001 | 0.001±0.012 | 0.004±0.014 | 0.004±0.014 | **0.000±0.012** | 0.002±0.012 | 0.004±0.014 |
| Unattended $PCC_2$ | | 0.003±0.001 | 0.002±0.012 | 0.003±0.013 | 0.004±0.013 | **0.000±0.012** | 0.002±0.012 | 0.003±0.013 |
| $\Delta PCC_1$ | | 0.015±0.001 | 0.012±0.018 | 0.013±0.019 | **0.016±0.020** | 0.009±0.018 | 0.011±0.018 | 0.013±0.019 |
| $\Delta PCC_2$ | | 0.016±0.002 | 0.012±0.018 | 0.014±0.020 | **0.017±0.021** | 0.009±0.018 | 0.010±0.018 | 0.014±0.020 |

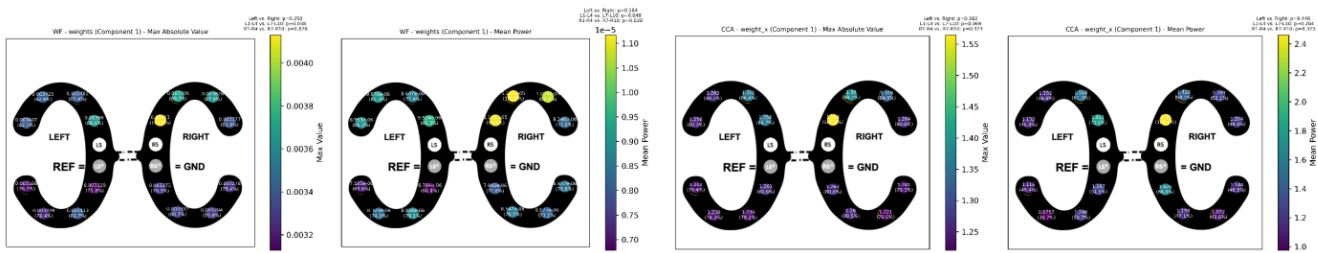

**Figure 3 Max and mean power of filter weights of each cEEGrid channels.** Left: weights of Wiener Filter (WF). Right: EEG weights of Canonical Component Analysis (CCA). In each channel shows the absolute and relative (percentage) weights averaged across all folds and subjects. T-test is applied to find out possible statistical difference between the left channels and the right channels, the upper and lower channels on the left and right ears. T-test results are demonstrated above the colormap.

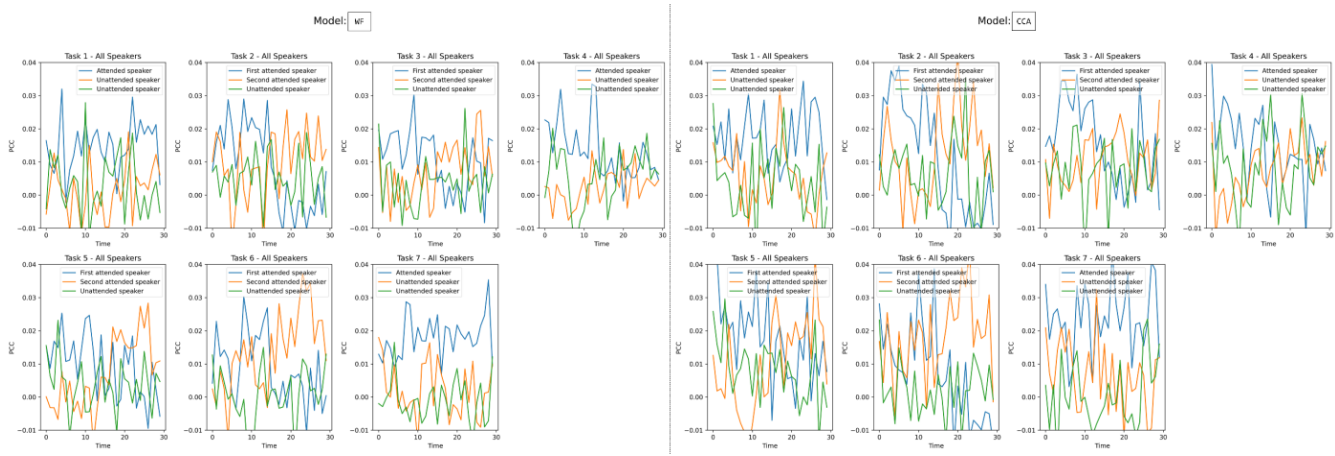

**Figure 4 Pearson's correlation coefficient (PCC) changes over time.** 30 seconds of reconstructed envelope is divided into non-overlapping 1 second segments and compared to the 1-second-long envelopes of the competing speakers. Task 2, 3, 4, 5, 6 involve attentional state switch at approximately the middle point of the trial (15 second), while task 1 and 7 do not involve any attentional switch. The PCC results are averaged across all subjects.

Accordingly, an additional experiment is conducted to evaluate the performance of the WF model across varying electrode configurations using a 30-second decision window. The results presented in Table 5 reveal that the WF model attained optimal decoding performance when eight electrodes were positioned on the upper portion of both ears (four electrodes on the left ear and four on the right ear), surpassing the full electrode configuration. In contrast, decoding performance degraded notably when only the eight lower electrodes were utilized. These findings suggest that the upper electrodes in the cEEGrid layout contain more informative neural signals for auditory attention decoding relative to the lower electrodes. The observed performance decline with the lower electrodes indicates that these channels carry less discriminative neural patterns.

Furthermore, results revealed that decoders exhibited modestly enhanced performance when utilizing electrodes positioned proximal to the right ear relative to those near the left ear. It is well-established that while both left and right temporal lobes contribute to auditory and language processing, the right temporal lobe plays a pivotal role in spatial localization processes []. Consequently, a plausible explanation is that in contexts with multiple competing speakers, the right temporal lobe becomes more highly engaged in identifying the target speaker and localizing the sound source, thereby generating more discriminative neural patterns for auditory attention decoding. Further investigation into the distinct functional contributions of the left and right temporal lobes in auditory attention decoding could generate novel insights and inform the development of streamlined daily-use auditory attention decoding devices.

*D. Tracking the attentional switch*

The Pearson correlation coefficient (PCC) are computed between the reconstructed envelope and the envelopes of competing speakers using a sliding window approach, with the resulting time-PCC plots depicted in Figure 4. These time-PCC curves were averaged across all participants and test folds. While fluctuations were observed across all experimental tasks, the attended PCC consistently remained higher than the unattended PCC values. Additionally, both the WF and CCA models demonstrated the ability to track shifts in attentional state (occurring around 15 seconds) in Tasks 2, 3, 4, 5, and 6. In contrast, for Tasks 1 and 7—where no attentional state changes were introduced—the attended PCC oscillated yet consistently remained above the unattended PCC values. The capacity of these linear models to track dynamic attentional shifts supports the potential utility of envelope reconstruction in monitoring listeners' attentional focus and its variations through cEEGrid signals.

Compared to Task 1, a more substantial deviation between attended and unattended PCC values was observed in Task 7. The task requirements (described in Table 1) elevated attentional demand, leading to heightened participant engagement and more distinct neural representations of the attended speaker. In Task 4, PCC exhibited a significant decrease following the switching point, where participants were instructed to disregard all speakers upon receiving a prompt. The time-PCC trajectory in Task 4 reflects this state of ignored attention, indicating that linear regression decoders can also detect absolute auditory attention decoding (sAAD)—specifically, they can distinguish whether a listener is attending to a particular sound source. However, it should be noted that decoding absolute AAD entails more than merely tracking temporal variations in PCC [58], and the implementation of such algorithms falls outside the scope of the present study.

Furthermore, no significant differences in decoding performance were observed across distinct task types. A plausible explanation is the limited size of the training dataset during task- and subject-specific envelope reconstruction. Additionally, certain tasks (e.g., Task 2 and Task 5) may involve similar types of attentional transitions, diminishing detectable performance differences between them.

## VI. CONCLUSIONS

This study introduces a novel auditory attention dataset comprising ear-EEG signals recorded from 98 participants using 16-channel cEEGrid electrodes. Participants were exposed to three concurrent speakers and engaged in tasks designed to capture multiple types of attentional transitions. Several linear and non-linear models were evaluated for their ability to decode either the attended envelope or the direction of attention from the ear-EEG signals. To mitigate overestimation of decoding performance, both leave-one-trial-out and nested leave-one-trial-out experimental protocols were implemented. The WF model achieved the highest decoding accuracy of $0.441 \pm 0.011$ under the nested leave-one-trial-out condition, with hyperparameters optimized using validation sets across all participants and a 30-second decision window. While models with individually tuned hyperparameters exhibited superior accuracy in leave-one-trial-out experiments, their performance declined in the more rigorous nested leave-one-trial-out settings. Both CSP and RGC classifiers outperformed WF and CCA models for decision window lengths shorter than 10 seconds but showed degraded performance with longer EEG segments. Further analysis corroborates the feasibility of tracking dynamic auditory attentional shifts using WF and CCA models applied to cEEGrid-recorded ear-EEG signals. Additionally, the four electrodes positioned on the upper portion of each ear's cEEGrid layout were found to provide more informative signals for auditory attention decoding compared to lower electrodes, with right-ear electrodes capturing more discriminative neural patterns than those on the left ear. Overall, these findings provide practical insights and establish a solid foundational basis for advancing research on ear-EEG-based auditory attention decoding, with implications for future brain–computer interface systems and real-world auditory attention applications. These results also stress the limitation of decoding attended envelope from multiple competing speakers with DNN models.

## REFERENCES

[1] E. C. Cherry, "Some experiments on the recognition of speech, with one and with two ears," *J. Acoust. Soc. Am.*, vol. 25, no. 5, pp. 975–979, 1953.

[2] A. W. Bronkhorst, "The cocktail-party problem revisited: early processing and selection of multi-talker speech," *Atten. Percept. Psychophys.*, vol. 77, no. 5, pp. 1465–1487, July 2015, doi: 10.3758/s13414-015-0882-9.

[3] I. Pollack and J. M. Pickett, "Cocktail party effect," *J. Acoust. Soc. Am.*, vol. 29, no. 11, pp. 1262–1262, Nov. 1957, doi: 10.1121/1.1919140.

[4] R. Plomp, "Auditory handicap of hearing impairment and the limited benefit of hearing aids," *J. Acoust. Soc. Am.*, vol. 63, no. 2, pp. 533–549, Feb. 1978, doi: 10.1121/1.381753.

[5] S. Vandecappelle, L. Deckers, N. Das, A. H. Ansari, A. Bertrand, and T. Francart, "EEG-based detection of the locus of auditory attention with convolutional neural networks," *eLife*, vol. 10, Apr. 2021, doi: 10.7554/eLife.56481.

[6] E. J. Ozmeral, "The effects of hearing impairment on the ability to glimpse speech in a spectro-temporally complex noise," Ph.D., The University of North Carolina at Chapel Hill, United States -- North Carolina, 2013. Accessed: May 07, 2025. [Online]. Available: https://www.proquest.com/docview/1492973412/abstract/2BD6B30B04564AF1PQ/1

[7] W. Biesmans, J. Vanthornhout, J. Wouters, M. Moonen, T. Francart, and A. Bertrand, "Comparison of speech envelope extraction methods for EEG-based auditory attention detection in a cocktail party scenario," in *2015 37th Annual International Conference of the IEEE Engineering in Medicine and Biology Society (EMBC)*, Milan: IEEE, Aug. 2015, pp. 5155–5158. doi: 10.1109/EMBC.2015.7319552.

[8] H. Candra *et al.*, "Investigation of window size in classification of EEG-emotion signal with wavelet entropy and support vector machine," in *2015 37th Annual International Conference of the IEEE Engineering in Medicine and Biology Society (EMBC)*, Aug. 2015, pp. 7250–7253. doi: 10.1109/EMBC.2015.7320065.

[9] A. de Cheveigné, D. D. E. Wong, G. M. Di Liberto, J. Hjortkjær, M. Slaney, and E. Lalor, "Decoding the auditory brain with canonical component analysis," *NeuroImage*, vol. 172, pp. 206–216, May 2018, doi: 10.1016/j.neuroimage.2018.01.033.

[10] J. Yang *et al.*, "Thoughts of brain EEG signal-to-text conversion using weighted feature fusion-based multiscale dilated adaptive DenseNet with attention mechanism," *Biomed. Signal Process. Control*, vol. 86, p. 105120, Sept. 2023, doi: 10.1016/j.bspc.2023.105120.

[11] X. Feng, X. Feng, B. Qin, and T. Liu, "Aligning Semantic in Brain and Language: A Curriculum Contrastive Method for Electroencephalography-to-Text Generation," *IEEE Trans. Neural Syst. Rehabil. Eng.*, pp. 1–1, 2023, doi: 10.1109/TNSRE.2023.3314642.

[12] E. Ceolini, I. Kiselev, and S.-C. Liu, "Evaluating Multi-Channel Multi-Device Speech Separation Algorithms in the Wild: A Hardware-Software Solution," *IEEEACM Trans. Audio Speech Lang. Process.*, vol. 28, pp. 1428–1439, 2020, doi: 10.1109/TASLP.2020.2989545.

[13] S. Geirnaert, T. Francart, and A. Bertrand, "Riemannian geometry-based decoding of the directional focus of auditory attention using EEG," in *ICASSP 2021 - 2021 IEEE International Conference on Acoustics, Speech and Signal Processing (ICASSP)*, June 2021, pp. 1115–1119. doi: 10.1109/ICASSP39728.2021.9413404.

[14] B. Holtze, M. Rosenkranz, M. Jaeger, S. Debener, and B. Mirkovic, "Ear-EEG Measures of Auditory Attention to Continuous Speech," *Front. Neurosci.*, vol. 16, 2022, Accessed: Oct. 30, 2023. [Online]. Available: https://www.frontiersin.org/articles/10.3389/fnins.2022.869426

[15] C. Fan *et al.*, "Seeing helps hearing: A multi-modal dataset and a mamba-based dual branch parallel network for auditory attention decoding," *Inf. Fusion*, vol. 118, p. 102946, June 2025, doi: 10.1016/j.inffus.2025.102946.

[16] B. Accou, J. Vanthornhout, H. V. Hamme, and T. Francart, "Decoding of the speech envelope from EEG using the VLAAI deep neural network," *Sci. Rep.*, vol. 13, no. 1, Art. no. 1, Jan. 2023, doi: 10.1038/s41598-022-27332-2.

[17] E. Alickovic, C. F. Mendoza, A. Segar, M. Sandsten, and M. A. Skoglund, "Decoding Auditory Attention From EEG Data Using Cepstral Analysis," in *2023 IEEE International Conference on Acoustics, Speech, and Signal Processing Workshops (ICASSPW)*, June 2023, pp. 1–5. doi: 10.1109/ICASSPW59220.2023.10193192.

[18] S. Geirnaert, T. Francart, and A. Bertrand, "Fast EEG-based decoding of the directional focus of auditory attention using common spatial patterns," *IEEE Trans. Biomed. Eng.*, pp. 1–1, 2020, doi: 10.1109/TBME.2020.3033446.

[19] Y. Zhang, H. Ruan, Z. Yuan, H. Du, X. Gao, and J. Lu, "A learnable spatial mapping for decoding the directional focus of auditory attention using EEG," in *ICASSP 2023 - 2023 IEEE International Conference on Acoustics, Speech and Signal Processing (ICASSP)*, June 2023, pp. 1–5. doi: 10.1109/ICASSP49357.2023.10096819.

[20] D. D. E. Wong, S. A. Fuglsang, J. Hjortkjær, E. Ceolini, M. Slaney, and A. de Cheveigné, "A comparison of regularization methods in forward and backward models for auditory attention decoding," *Front. Neurosci.*, vol. 12, 2018, doi: 10.3389/fnins.2018.00531.

[21] S. Cai, P. Li, and H. Li, "A Bio-Inspired Spiking Attentional Neural Network for Attentional Selection in the Listening Brain," *IEEE Trans. Neural Netw. Learn. Syst.*, pp. 1–11, 2023, doi: 10.1109/TNNLS.2023.3303308.

[22] S. Cai, R. Zhang, and H. Li, "Robust Decoding of the Auditory Attention from EEG Recordings Through Graph Convolutional Networks," in *ICASSP 2024 - 2024 IEEE International Conference on Acoustics, Speech and Signal Processing (ICASSP)*, Seoul, Korea, Republic of: IEEE, Apr. 2024, pp. 2320–2324. doi: 10.1109/ICASSP48485.2024.10447633.

[23] C. Fan *et al.*, "DGSD: Dynamical graph self-distillation for EEG-based auditory spatial attention detection," *Neural Netw.*, vol. 179, p. 106580, Nov. 2024, doi: 10.1016/j.neunet.2024.106580.

[24] Y. Ding *et al.*, "EEG-Deformer: A Dense Convolutional Transformer for Brain-Computer Interfaces," *IEEE J. Biomed. Health Inform.*, vol. 29, no. 3, pp. 1909–1918, Mar. 2025, doi: 10.1109/JBHI.2024.3504604.

[25] Y. Zhang, J. Lu, F. Chen, H. Du, X. Gao, and Z. Lin, "Multi-Class Decoding of Attended Speaker Direction

Using Electroencephalogram and Audio Spatial Spectrum," *IEEE Trans. Neural Syst. Rehabil. Eng.*, vol. 33, pp. 2892–2903, 2025, doi: 10.1109/TNSRE.2025.3591819.

[26] C. Fan *et al.*, "M3ANet: Multi-scale and Multi-Modal Alignment Network for Brain-Assisted Target Speaker Extraction," May 31, 2025, *arXiv*: arXiv:2506.00466. doi: 10.48550/arXiv.2506.00466.

[27] C. Fan, S. Zhang, J. Zhang, Z. Pan, and Z. Lv, "SSM2Mel: State Space Model to Reconstruct Mel Spectrogram from the EEG," in *ICASSP 2025 - 2025 IEEE International Conference on Acoustics, Speech and Signal Processing (ICASSP)*, Apr. 2025, pp. 1–5. doi: 10.1109/ICASSP49660.2025.10888785.

[28] A. Gu and T. Dao, "Mamba: Linear-Time Sequence Modeling with Selective State Spaces," May 31, 2024, *arXiv*: arXiv:2312.00752. doi: 10.48550/arXiv.2312.00752.

[29] T. Dao and A. Gu, "Transformers are SSMs: Generalized Models and Efficient Algorithms Through Structured State Space Duality," May 31, 2024, *arXiv*: arXiv:2405.21060. doi: 10.48550/arXiv.2405.21060.

[30] H. Chao, Y. Cao, and Y. Liu, "Multi-channel EEG emotion recognition through residual graph attention neural network," *Front. Neurosci.*, vol. 17, p. 1135850, July 2023, doi: 10.3389/fnins.2023.1135850.

[31] T. Song, W. Zheng, P. Song, and Z. Cui, "EEG emotion recognition using dynamical graph convolutional neural networks," *IEEE Trans. Affect. Comput.*, vol. 11, no. 3, pp. 532–541, July 2020, doi: 10.1109/TAFFC.2018.2817622.

[32] M. G. Bleichner, B. Mirkovic, and S. Debener, "Identifying auditory attention with ear-EEG: cEEGrid versus high-density cap-EEG comparison," *J. Neural Eng.*, vol. 13, no. 6, p. 066004, Oct. 2016, doi: 10.1088/1741-2560/13/6/066004.

[33] H. Zhu *et al.*, "Using Ear-EEG to Decode Auditory Attention in Multiple-speaker Environment," in *ICASSP 2025 - 2025 IEEE International Conference on Acoustics, Speech and Signal Processing (ICASSP)*, Apr. 2025, pp. 1–5. doi: 10.1109/ICASSP49660.2025.10890810.

[34] L. Fiedler, M. Wöstmann, C. Graversen, A. Brandmeyer, T. Lunner, and J. Obleser, "Single-channel in-ear-EEG detects the focus of auditory attention to concurrent tone streams and mixed speech," *J. Neural Eng.*, vol. 14, no. 3, p. 036020, Apr. 2017, doi: 10.1088/1741-2552/aa66dd.

[35] E. Larson and A. K. C. Lee, "The cortical dynamics underlying effective switching of auditory spatial attention," *NeuroImage*, vol. 64, pp. 365–370, Jan. 2013, doi: 10.1016/j.neuroimage.2012.09.006.

[36] W. Pu, P. Zan, J. Xiao, T. Zhang, and Z.-Q. Luo, "Evaluation of joint auditory attention decoding and adaptive binaural beamforming approach for hearing devices with attention switching," in *ICASSP 2020 - 2020 IEEE International Conference on Acoustics, Speech and Signal Processing (ICASSP)*, May 2020, pp. 8728–8732. doi: 10.1109/ICASSP40776.2020.9054592.

[37] S. Haro, H. M. Rao, T. F. Quatieri, and C. J. Smalt, "EEG alpha and pupil diameter reflect endogenous auditory attention switching and listening effort," *Eur. J. Neurosci.*, vol. 55, no. 5, pp. 1262–1277, 2022, doi: 10.1111/ejn.15616.

[38] I. Rotaru, S. Geirnaert, N. Heintz, I. V. de Ryck, A. Bertrand, and T. Francart, "What are we really decoding? Unveiling biases in EEG-based decoding of the spatial focus of auditory attention," *J. Neural Eng.*, vol. 21, no. 1, p. 016017, Feb. 2024, doi: 10.1088/1741-2552/ad2214.

[39] S. Cai, P. Sun, T. Schultz, and H. Li, "Low-latency auditory spatial attention detection based on spectro-spatial features from EEG," in *2021 43rd Annual International Conference of the IEEE Engineering in Medicine Biology Society (EMBC)*, Nov. 2021, pp. 5812–5815. doi: 10.1109/EMBC46164.2021.9630902.

[40] S. Cai, E. Su, L. Xie, and H. Li, "EEG-based auditory attention detection via frequency and channel neural attention," *IEEE Trans. Hum.-Mach. Syst.*, pp. 1–11, 2021, doi: 10.1109/THMS.2021.3125283.

[41] S. Cai, T. Schultz, and H. Li, "Brain Topology Modeling With EEG-Graphs for Auditory Spatial Attention Detection," *IEEE Trans. Biomed. Eng.*, pp. 1–11, 2023, doi: 10.1109/TBME.2023.3294242.

[42] C. Fan *et al.*, "ListenNet: A Lightweight Spatio-Temporal Enhancement Nested Network for Auditory Attention Detection," May 15, 2025, *arXiv*: arXiv:2505.10348. doi: 10.48550/arXiv.2505.10348.

[43] I. Rotaru, S. Geirnaert, N. Heintz, I. Van De Ryck, A. Bertrand, and T. Francart, "EEG-based decoding of the spatial focus of auditory attention in a multi-talker audiovisual experiment using Common Spatial Patterns," Neuroscience, preprint, July 2023. doi: 10.1101/2023.07.13.548824.

[44] Y. Jiang, N. Chen, and J. Jin, "Detecting the locus of auditory attention based on the spectro-spatial-temporal analysis of EEG," *J. Neural Eng.*, vol. 19, no. 5, p. 056035, Oct. 2022, doi: 10.1088/1741-2552/ac975c.

[45] Y. Yan *et al.*, "Auditory Attention Decoding in Four-Talker Environment with EEG," in *Interspeech 2024*, ISCA, Sept. 2024, pp. 432–436. doi: 10.21437/Interspeech.2024-739.

[46] Y. Yan *et al.*, "Overestimated performance of auditory attention decoding caused by experimental design in EEG recordings," presented at the Proc. Interspeech 2025, 2025, pp. 1053–1057. doi: 10.21437/Interspeech.2025-85.

[47] Z. Yuanming, S. Zeyan, L. Jing, and L. Zhibin, "16 channel Three speaker dynamic switch cEEGrid Auditory Attention Decoding Dataset Nanjing University." Zenodo, Oct. 20, 2025. doi: 10.5281/zenodo.17393865.

[48] Yuanming Zhang, Zeyan Song, Jing Lu, and Zhibin Lin, "16 channel Three speaker dynamic switch cEEGrid Auditory Attention Decoding Dataset Nanjing University." IEEE DataPort. doi: 10.21227/7QPK-9J22.

[49] S. Blum, N. S. J. Jacobsen, M. G. Bleichner, and S. Debener, "A Riemannian Modification of Artifact

Subspace Reconstruction for EEG Artifact Handling," *Front. Hum. Neurosci.*, vol. 13, Apr. 2019, doi: 10.3389/fnhum.2019.00141.

[50] M. Plechawska-Wojcik, M. Kaczorowska, and D. Zapala, "The Artifact Subspace Reconstruction (ASR) for EEG Signal Correction. A Comparative Study," in *Information Systems Architecture and Technology: Proceedings of 39th International Conference on Information Systems Architecture and Technology – ISAT 2018*, J. Świątek, L. Borzemski, and Z. Wilimowska, Eds., Cham: Springer International Publishing, 2019, pp. 125–135. doi: 10.1007/978-3-319-99996-8_12.

[51] L. Pion-Tonachini, K. Kreutz-Delgado, and S. Makeig, "ICLabel: An automated electroencephalographic independent component classifier, dataset, and website," *NeuroImage*, vol. 198, pp. 181–197, Sept. 2019, doi: 10.1016/j.neuroimage.2019.05.026.

[52] T. Akiba, S. Sano, T. Yanase, T. Ohta, and M. Koyama, "Optuna: A Next-generation Hyperparameter Optimization Framework," in *Proceedings of the 25th ACM SIGKDD International Conference on Knowledge Discovery & Data Mining*, in KDD '19. New York, NY, USA: Association for Computing Machinery, July 2019, pp. 2623–2631. doi: 10.1145/3292500.3330701.

[53] hvy, "'Multivariate' TPE Makes Optuna Even More Powerful," Optuna. Accessed: Sept. 01, 2025. [Online]. Available: https://medium.com/optuna/multivariate-tpe-makes-optuna-even-more-powerful-63c4bfbaebe2

[54] S. Watanabe, "Significant Speed Up of Multi-Objective TPESampler in Optuna v4.0.0," Optuna. Accessed: Sept. 01, 2025. [Online]. Available: https://medium.com/optuna/significant-speed-up-of-multi-objective-tpesampler-in-optuna-v4-0-0-2bacdcd1d99b

[55] S. Watanabe, "Tree-Structured Parzen Estimator: Understanding Its Algorithm Components and Their Roles for Better Empirical Performance," May 26, 2023, *arXiv*: arXiv:2304.11127. doi: 10.48550/arXiv.2304.11127.

[56] J. Bergstra, D. Yamins, and D. D. Cox, "Making a Science of Model Search: Hyperparameter Optimization in Hundreds of Dimensions for Vision Architectures".

[57] I. Loshchilov and F. Hutter, "Decoupled Weight Decay Regularization," in *7th International Conference on Learning Representations, ICLR 2019, New Orleans, LA, USA, May 6-9, 2019*, OpenReview.net, 2019. Accessed: Oct. 16, 2025. [Online]. Available: https://openreview.net/forum?id=Bkg6RiCqY7

[58] A. Roebben, N. Heintz, S. Geirnaert, T. Francart, and A. Bertrand, "'Are you even listening?' - EEG-based decoding of absolute auditory attention to natural speech," *J. Neural Eng.*, 2024, doi: 10.1088/1741-2552/ad5403.